\documentclass[twocolumn,showpacs,preprintnumbers,amsmath,amssymb,pre]{revtex4}
\usepackage{graphicx}
\usepackage{graphics}
\usepackage{subfigure}
\usepackage{dcolumn}
\usepackage{bm}
\usepackage{txfonts}

\begin{document}

\preprint{}

\title{Partner selections in public goods games with constant group size}

\author{Te Wu$^1$}
\author{Feng Fu $^{1,2}$}
\author{Long Wang$^1$}
\email{longwang@pku.edu.cn}
\affiliation{ $^1$Center for Systems and
Control, State Key Laboratory for Turbulence and Complex Systems,
College of Engineering, Peking University, Beijing 100871, China\\
$^2$Program for Evolutionary Dynamics, Harvard University,
Cambridge, Massachusetts 02138, USA}

\date{\today}
\begin{abstract}
Most of previous studies concerning the Public Goods Game assume
either participation is unconditional or the number of actual
participants in a competitive group changes over time. How the fixed
group size, prescribed by social institutions, affects the evolution
of cooperation is still unclear. We propose a model where
individuals with heterogeneous social ties might well engage in
differing numbers of Public Goods Games, yet with each Public Goods
Game being constant size during the course of evolution. To do this,
we assume that each focal individual unidirectionally selects a
constant number of interaction partners from his immediate neighbors
with probabilities proportional to the degrees or the reputations of
these neighbors, corresponding to degree-based partner selection or
reputation-based partner selection, respectively. Because of the
stochasticity the group formation is dynamical. In both selection
regimes, monotonical dependence of the stationary density of
cooperators on the group size was found, the former over the whole
range but the latter over a restricted range of the renormalized
enhancement factor. Moreover, the reputation-based regime can
substantially improve cooperation. To interpret these differences,
the microscopic characteristics of individuals are probed. We later
extend the degree-based partner selection to general cases where
focal individuals have preferences towards their neighbors of
varying social ties to form groups. As a comparison, we as well
investigate the situation where individuals locating on the degree
regular graphs choose their co-players at random. Our results may
give some insights into better understanding the widespread teamwork
and cooperation in the real world.
\end{abstract}
\pacs{89.75.Hc, 87.23.Kg, 02.50.Le}

 \keywords{}
 \maketitle
\section{Introduction}
As a most popular game, the Prisoner's Dilemma (PD) has been widely
employed to characterize and elucidate the cooperation conundrum
between individual and group interests through pairwise
interactions~\cite{Axelrod1, Axelrod2, Traulsen, Vukov05pre,
Vukov06pre, Zimmermann05pre, Zhi-Xi Wu06pre, McNamara04n}. In many
realistic situations ranging from cellular organisms to hunter
business to national negotiations, however, multiple agents instead
of two individuals are usually involved. While many researchers
treated these $N$-person problems as a summation of many two-person
problems~\cite{Axelrod1, Axelrod2, Wang06pre, Ohtsuki06n, Hofbauer,
Hauert04N}, the Public Goods Game (PGG) was proposed as a
representative of built-in interactions to investigate the
multi-person predicament of cooperation, which can be regarded as a
natural extension of Prisoner's Dilemma~\cite{Hauert02s, Szabo02Prl,
Semmann03N}. In a typical example of the PGG, players belonging to a
community of $N$ individuals can adopt one of the feasible actions,
say cooperation (C) and defection (D). A cooperator donates an
amount of $c$ investment to the common pool whereas a defector
nothing. The sum is augmented by an enhancement factor $r$ and then
equally distributed among all players irrespective of their
contributions~\cite{Hauert02s}. In accordance with the name of the
PGG, the parameter $r$ should be constrained to be less than the
group size of the PGG but larger than unit (i.e., $1<r<N$),
suggesting that group of cooperators are better off than group of
defectors whereas defectors outperform cooperators in any given
mixed group~\cite{Doebeli04s}. Despite that the group end up
maximizing their payoff if all cooperate, the best strategy for a
player is to defect, since every invested unit contribution is
discounted as a return~\cite{Hauert02jtb}. Thus, the social dilemma
of what is best for egoistic individual and what is best for the
group arises. According to both classic and evolutionary game
theory, cooperators are doomed under natural selection, which is
usually at odds with the observations in the real world.
Considerable efforts have been expended to find solutions to this
plausible paradox.

A variety of measures, such as punishment~\cite{Clutton95N,
Dreber08N, Christoph07S, Reckonbach06N, Fehr02N}, social diversity
and the associated diversity of contribution~\cite{Santos08N,
Perc08pre, Doebeli04s}, optional participation~\cite{Hauert02s,
Szabo02Prl, Hauert02jtb} and image score effect~\cite{Brandt05pnas,
FuFeng08pre, Wedekind0s}, have been proposed to answer the question
how large-scale cooperation can evolve and persist stably. In
Ref.~\cite{Santos08N}, the authors investigated the influence of two
different patterns of contribution on the cooperators' evolutionary
fate in the context of the PGG whenever individuals interact along
the heterogeneous social ties, and concluded that cooperation can be
enhanced if any act of giving is considered to be cooperative,
irrespective of the amount of giving. It should be noted that in
this work individuals play PGGs with all those directly connected to
them, naturally introducing coercion of participation. Different
from this assumption, the autarkic 'loner' was introduced as a third
strategy besides cooperation and defection in
Refs.~\cite{Hauert02jtb, Hauert02s, Szabo02Prl}. It has shown that
this voluntary participation efficiently prevents defectors from
spreading within the population through self-adjusting the group
size of the PGG, leading to the appearance of the cyclic dominance
of Rock-Scissor-Paper type. In most such investigations,
nevertheless, the effective group size varies over time as the
frequencies of these three strategies oscillate in the population.

In most ubiquitous observed public goods type
interactions~\cite{Binmore, Dugatkin, Colman}, however, the group
formation is not always in this way. Due to the restriction of some
social norms and, the fact that each individual has the right to
decide whether or not to attend an activity~\cite{Lee08pa}, the
group size can neither be arbitrary number nor be equal to the
number of one's neighbors. Instead, individuals are usually divided
into equal subgroups to accomplish a public target: of these
examples are student dormitory clean, public transportation and
predator inspection behavior. On the one hand, although large teams,
clubs can function most effectively if their members get well along
one another, some individuals are easily tempted to free-ride on the
public resources without incurring any cost of contribution~\cite{
Coricelli04j}. On the other hand, it is difficult to accomplish a
public task if too few persons engage in it~\cite{Haag07jet}. Thus,
one may ask what the invariable group size of the PGG should be, and
how it influences the survivability of
cooperators~\cite{Manfred06pnas, Manfred08pnas, Breteville07el}.

In this paper we set up a minimal model in which individuals of
varying social ties maybe participate different numbers of PGGs
while the group size of each typical PGG remains constant during the
evolutionary process, in line with most already performed public
goods experiments where samples (\emph{usually students}) were
actually divided into groups containing equal rather than
heterogeneous members~\cite{Manfred06pnas}. The Barab\'{a}si-Albert
scale-free networks are adopted to represent interpersonal
connections, since most natural and artificial networks share much
in common with this type of networks~\cite{Amaral0pnas}. Focal
individuals take into account two regimes to select the group
members when playing PGGs: degree-based and reputation-based partner
selections. In the partner selection based on degree, a focal
individual selects his $g$ neighbors with probabilities concerned
with their social ties. For the convenience of discussion, we divide
the enhancement factor $r$ by the group size $g+1$ to be the
renormalized enhancement factor $\delta$. By virtue of numerical
simulations, we found that for small group size, the system
transforms from one homogeneous state of defectors, after a sharp
transition, into the uniform state of cooperators for increasing
$\delta$. For large group size, the curve of cooperation lever
versus the quantity $\delta$ sees a 'gentle slope', implying that
cooperators and defectors can coexist for a wide spectrum of
$\delta$. In the partner selection based on reputation, a focal
individual chooses his co-players among his neighbors with
probabilities associated with their reputations. Intriguingly,
cooperation level as a function of $\delta$ monotonically increases
almost parallelly as $\delta$ increases responding to different
group sizes. Thus, the interplay of group size and selection regime
together orients the evolution and accordingly leading to disparate
dynamics of the population. Besides, we extend the degree-based
partner selection by equipping focal individuals with biases towards
their neighbors of different social ties. As a comparison, we also
perform the numerical simulation on degree regular graphs.

The rest of this paper is structured as follows. We make a brief
introduction of our model in Section II. Numerical results as well
as discussions to these accomplished results are presented in
Section III. Concluding remarks are drawn in Section IV.

\section{Model}
We consider a system with constant population size $N$. The pairwise
connections are specified via a Barab\'{a}si-Albert network, in
which each vertex represents an individual. To construct such a
network, we start from a small ring evenly embedded with $m_0$
nodes. At each time step, a newly added node links to $m$ existing
nodes in the instantaneous network following the preferential
attachment scheme, $\emph{i.e. }$, the probability of an existing
node attracting a link is proportional to its current
degree~\cite{barabasi98s}. We repeat this process until $N$ nodes
are present in the network. Initially, half proportion of the
population are randomly assigned to be cooperators (C) and the
remaining defectors (D). Instead of assuming compulsory
participation in the PGG~\cite{Santos08N}, each individual (focal
individual) picks out a fixed number $(g)$ of his neighbors to join
in the public enterprize according to the specified partner
selection regime. Due to the heterogeneity of connections, different
individuals can potentially involve in diverse numbers of PGGs and,
the diversity in the numbers is closely associated with individuals'
social ties or reputations. The payoff of a certain individual is
accumulated over the sum of all the PGGs centered on his neighbors
and himself, respectively. Following common practice, individual
obtaining higher payoff are more likely to disseminate his strategy.
After each round of the game, each individual $i$ compares his
payoff ($P_i$) with that ($P_j$) of a randomly chosen neighbor $j$
and switches his strategy $s_i$ to $s_j$ with a probability
$T(s_i\rightarrow s_j)=(P_j-P_i)/M$ whenever j fares better
(provided the payoff difference is positive), with $M$ ensuring the
proper normalization and being given by the maximal possible
difference between payoffs of $i$ and $j$. Otherwise, he maintains
his present strategy.

As for selection regime, two different patterns are considered here.
First, given that individuals of varying numbers of social ties play
distinctly different role in real-world communities, thus, whenever
playing the PGG, a focal individual has the privilege to determine
which neighbors to be picked up. For a specific PGG centered on
individual $i$, the probability that each of his neighbors is chosen
is given by \begin{equation}
Q(j)=\frac{k_j^\beta}{\Sigma_{j\in\Omega_i} k_j^\beta}
\end{equation} where $k_j$ is the number of neighbors of $i$'s
$j$\emph{th} neighbor and $\Omega_i$ the neighborhood set of
individual $i$. Apparently, the exponent $\beta$ which we define as
the weight of participation, uniquely measures to which extent this
partner selection is related to degree. In other words, $\beta=0$
represents that all individuals though with heterogeneous social
ties have the same opportunity to be selected, indicating that focal
individuals view their neighbors indiscriminately. Highly connected
individuals behave actively whenever $\beta$ is positive and
otherwise corresponds to the opposite situation. Individual picks up
deterministically his most connected neighbor and second most
connected and so on when the parameter $\beta$ takes the value of
infinity.

Besides, in repeated games, rational individuals can acquire
information of their neighbors' performance during the past moves,
which ineluctably has influence upon individual decision making of
either continuing to play with these neighbors or replace them with
alternative ones, if they exist, in the future rounds. In order for
maximizing one's own self-interest, individual tends to team up with
his more collaborative neighbors and interact in the Public Goods
Game. We, therefore, conceptualize an individual's times of
cooperation in the history as his reputation, known to all his
neighbors. Explicitly, $R_i(t+1)=R_i(t)+\delta_t$\, where $R_i(t)$
is individual $i$'s reputation at time step $t$. The function
$\delta_t$ takes value of unit if individual plays cooperation(C)
with his partners at time step $t$ and zero if he defects (for
details see~\cite{FuFeng08pre} and references thereof). To
ameliorate one's own income, a focal individual is inclined to enter
partnership with those who frequently cooperate, which means
individuals with larger reputations are more likely to engage in
more PGGs. Herein, we assume
\begin{equation}S(j)=\frac{R_j(t)}{\Sigma_{k\in\Omega_i}
R_k(t)}\end{equation} where $S(j)$ has the same definition as $Q(j)$
in the formula $(1)$ aforementioned. Evidently, focal individuals
measure the cooperativeness of their neighbors based on their
long-term performance other than decisions on one shot. A neighbor
defects once because of errors or other uncertainties, will not
affect his reputation greatly if he immediately retrieves
cooperation in next rounds. But if he frequently defect, focal
individual would regard him to be a bad one. Or rather, individuals
who defect once should not impose great influence upon themselves
but these sticking to defection would certainly suffer from being
excluded.

To decouple the effect imposed by the heterogenous social ties of
individuals situated on the Barab\'{a}si-Albert network from the
effect arising from selection regime, we also carried on our
simulation on degree regular graphs, creating from a ring with the
nearest- and next-nearest neighbors~\cite{szabo07pr}. To make a
clear comparison, the two types of homogeneous and inhomogeneous
networks have equivalent average connectivity.

\section{Results and discussion}

In our simulations, a population of $N$ individuals is considered
and a Barab\'{a}si-Albert network is generated to specify the
connections between them, along which interactions can occur.
Initially, equal percentage of cooperators and defectors are
randomly distributed among these $N$ nodes. Individuals spread their
strategies under replicate dynamics, meaning that the more fit
individuals are more likely to subsist and proliferate their
strategies within the population. In view that both individual's
social ties and reputation can affect one's social popularity, focal
individual tends to select his partners according to one of the two
features of his neighbors in our model. The synchronous update is
adopted for strategy evolution. We shall investigate how these two
different selection regimes separately affect the evolutionary fate
of cooperators when varying the interacting group size of the Public
Goods Game.

Let us first consider the situation of the weight of participation,
$\beta$, being zero, wherein each individual is picked up
equiprobably when playing PGGs, independent of whatever their social
ties are. The fraction of cooperators who survive at equilibrium
state is used to measure the evolution of cooperation. We plot the
fraction of these cooperators as a function of the renormalized
enhancement factor $\delta$ in Fig. 1. Obviously, as the PGG group
size contracts, the threshold ($\delta_{threshold}$) required above
which defectors are unable to wipe out cooperators anymore
appreciates. For a given $\delta(< 0.87)$, increasing the group size
leads to an enhancement of cooperation level. The monotonical
dependence becomes vague whenever $\delta$ approaches approximately
unit, but cooperation level $(\geq 0.75)$ has been high with respect
to the mean value $(0.5)$. As a whole, we can say that large group
size readily contributes to the emergence of cooperation in
comparison to small ones if it is invariable during the course of
evolution. Intuitively, the rationale behind this phenomenon is that
increasing group size offers the cooperative individuals with more
opportunities to meet more cooperators, thus leading to the
formation of cooperative clusters, which can resist replacement of
defectors successfully. Consider a cooperator surrounded by three
defector and two cooperator neighbors. His expected payoff would be
$1.4\delta-1$ if $g=1$ and $1.8\delta-1$ if $g=2$. Similarly,
defectors can also increase their payoff through exploiting more
cooperators per generation as cooperators do for larger group. The
benefit resulting from the aggrandized group acts on cooperators and
defectors' viability at different rates, the former stronger than
the latter, thereby lower quantity of $\delta$ can preferentially
select cooperators over defectors, as can be seen in Fig. 1.
Interestingly, disparities were found between our findings and the
results reported in Ref.~\cite{Santos08N}, in both scenarios with
focal individual contributing a fixed amount, $c$, to each PGG but
different numbers of possible PGGs he engages in. In our model, the
diversity in the group size vanishes. Thus, the heterogeneity in the
numbers of the PGGs each individual participates is annealed as
compared to the situation of compulsory participation. Notably, this
coercion in individual's playing games results in the interactions
of individuals as diverse as the population connections. The
diversity of fitness deduced from both types of above diversity
shrinks in our model, leading essentially to that cooperation level
as a function of $\delta$ observes a mild slope compared with
results in Ref.~\cite{Santos08N} for identical conditions with
exception of whether or not participation is obligatory. It is
noting that the PGG is reduced to the Prisoner's Dilemma for the
case of $g=1$, indicating focal individual plus one randomly
selected neighbor forms an interacting competed community, the
evolutionary equilibrium of the population switches from a uniform
state of defectors, after a sharp jump corresponding to a
coexistence state between defectors and cooperators, into a
homogeneous state of cooperators as $\delta$ increases.
\begin{figure}
\includegraphics[width=\columnwidth]{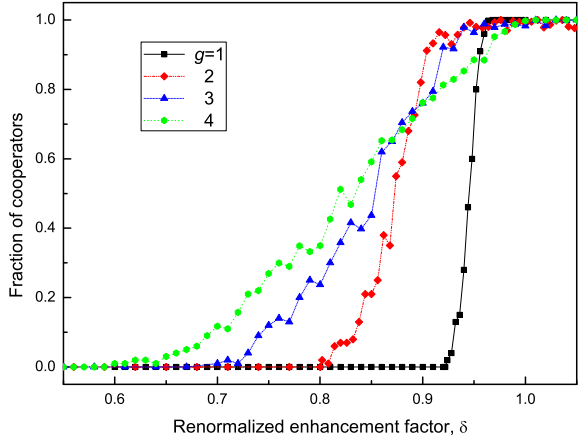}
\caption{(Color online) Evolution of cooperation. Fraction of
cooperators as a function of the renormalized enhancement factor
$\delta$, for different group sizes. Denoting the group size by
$g+1$. Population structure is characterized by heterogeneous graph
of $N=500$ nodes. We construct the graph following the growth and
preference attachment mechanism. Two associated parameters are
$m_0=4$ and $m=4$. Each data point is averaged over 100 runs, with
10 independent initial strategy distributions in each run.
}\label{fig1}
\end{figure}

Intriguingly, the monotonous dependence of the cooperative behaviors
on the constant group size turns inconspicuous when cooperators
greatly dominate the population (see Fig. 1). An inspection of the
microscopic property of the evolutionary process is indispensable to
understand the appearance of the nontrivial cooperation level versus
the group size. Fig. 2 shows the cooperator distribution among
individuals with different social ties in the population. For
$\delta\rightarrow \delta_{threhold}$ responding to lower
cooperation level, the frequency of cooperative strategies declines
from high-degree individuals to medium-degree to low-degree ones,
consistent with claims made in most previous relevant
works~\cite{Santos08N, rongzhihai08pre}. This positive correlation
is still existent for medium-value of $\delta$ wherein cooperators
and defectors are roughly equally distributed, but grows not so
apparent. It is not, however, the same case whenever cooperators
dominate defectors whereas unable to homogenize the whole
population. The social dilemma in the PGG is as strong as in the
Prisoner's Dilemma for small $\delta$. In this settings can
cooperators withstand the exploitation of defectors by forming the
cluster of compact uniform cooperative community, robust against
invasion of egoists through mutual breeding. As $\delta$ rises, the
dilemma is gradually lightened. The relaxation is especially evident
for $\delta$ approximating $1$. Our results approbate this
prediction (see Fig. 2). The decisive role hubs play in navigating
the evolutionary direction is weakened as opposed to the medium and
low-value of $\delta$, and consequently some hubs switch between
cooperation and defection from time to time (see the upper panel in
Fig. 2).
\begin{figure}
\includegraphics[width=\columnwidth]{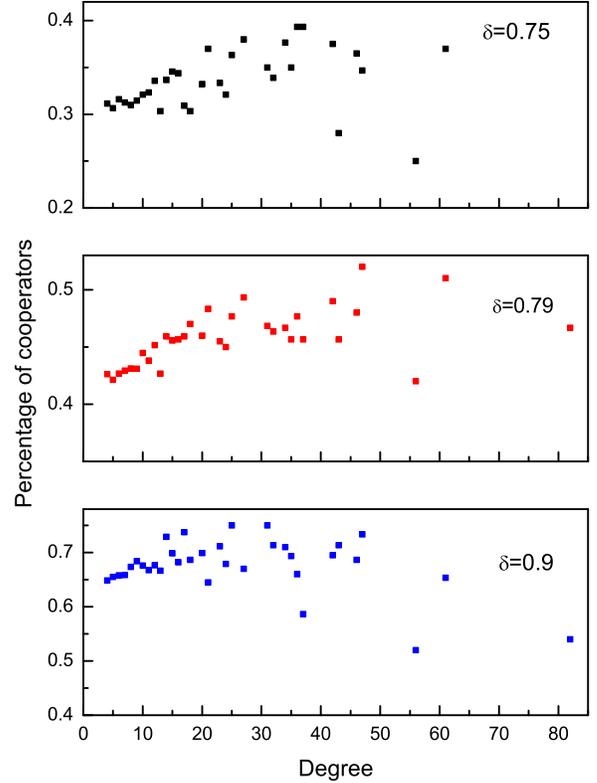}
\caption{(Color online) Fraction of cooperators as a function of
degree. For a fixed network, each data point is averaged over 300
runs with independent initial strategy distributions. Each run
progresses a time of 5000 rounds as transition, and a time window of
the next 1000 rounds is intercepted to collect data. A typical value
of $g$ is set to be 4. Cooperation level is 0.27, 0.51, 0.76 from
the upper panel to the below one, respectively.}\label{fig2}
\end{figure}

Unavoidably, rational individuals have capability of memorizing the
past performance of their neighbors if the game is repeatedly
played. In what follows, we consider another type of selection
regime, $\emph{i.e.}$, reputation-based partner selection. In this
regime, focal individuals decide which neighbors to interact with
based on the reputations of these neighbors. Fig. 3 demonstrates the
associated results. To reach the same nonzero cooperation level, the
larger group size, the lower $\delta$ is needed. Interestingly,
during the transient phases, cooperation level almost parallelly
increases with respect to $\delta$ for varying $g$, greatly
different from in the case where each individual is selected with
equal probability. In the degree-based partner selection, once the
social heterogeneous graph describing the population structure is
constructed, it would always be static. As a result of the selection
just pertaining to individual's social ties, the number of the PGGs
each takes part in is unchanged from the perspective of statistics,
independent of how often it cooperated in the past. Conversely, a
focal individual would alienate his such neighbors with lower
reputations by depressing the pick probabilities assigned to them,
on the condition he can get access to the local information bearing
on the history experience of his neighbors. Individual should be
cautious to make his strategy, therefore. For convenience of
simulation tractability, each individual was equally initialized
with a decimal reputation. If an individual frequently free-rides on
the public goods, his reputation progressively becomes compromised
relative to mostly cooperative individuals. Apparently, an
individual with a low reputation would be ostracized by the
communities centered on his neighbors, with an extreme case that an
individual of always defecting interpolating in a sea of cooperators
is destined to die since no individuals are willing to play with him
if other alternative neighbors can be found. Herein, this can be
thought as a positive feedback correlated the associated members in
a group with their reputations. Defectors disrepute themselves as
time goes, inducing the repulsion of their neighbors to continue to
team with them, which in turn becomes a controlling factor for
defectors to be in many PGGs and naturally reduces the opportunities
for them to exploit more cooperators. But for cooperators, the
inverse holds: focal individuals use the cooperativeness of these
cooperators as a choice criterion can help maintain these
cooperators a good standing, and thus will attract more neighbors to
choose them in future rounds. The exclusion of defectors by their
neighbors and the positive assortment among cooperative individuals
together brings that configuration fraught with defectors is
deteriorating, and will be relegated to cooperate, whereas the ones
full of cooperators will reinforce themselves and strengthen their
resistance against intrusion of defectors. Thus, impressively lower
$\delta$ can induce the emergence and maintenance of cooperation.
This reinforcement especially prevents the vacillation of some hubs
between cooperation and cooperation as appeared in degree-based
partner selection (see Fig. 2 and Fig. 4). Defectors are unable to
invade the evolving formed cooperative associates once fully
established as the reciprocal altruism of the assortment operates.

\begin{figure}
\includegraphics[width=\columnwidth]{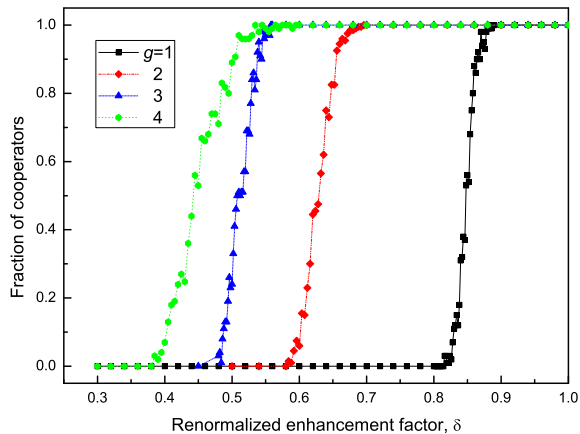}
\caption{(Color online) Fraction of cooperators as a function of the
parameter $\delta$. Focal individuals each select their group
members based on the reputation of their neighbors. Population
structure adopted is similar to that in Fig. 1. We have collected
the data points as in Fig. 1.}\label{fig3}
\end{figure}

We next extend the degree-based partner selection regime to more
generous cases by setting the weight of participation, $\beta$, to
be a series of discrete values (i.e., $\beta=-1, 0, 1,\infty $).
Here $\beta=\infty$ means that a focal individual chooses his
most-large and next-most-large neighbors as protagonists if $g=2$,
and the like. Results were illustrated in Fig. 5. One can find large
$\beta$ favors cooperation for constant group size. This observation
can be attributed to the effect of the parameter, $\beta$, on the
heterogeneous numbers of the PGGs each individual participates.
Although, the diversity of connections per individual are
unchangeable as we adopted a static network of contacts, our
assumption of constant group size makes the heterogeneity of network
not exactly coincide with that of the numbers of the PGGs each
individual actually engages in. Depending on the values of $\beta$,
the deviation can be either intensified or weakened. Condition
$\beta<0$ indicates individuals are preferential to interact with
low-degree ones among their neighbors, which favors unsociable
individuals to involve more games while opposes gregarious
individuals, who are surrounded with packed neighbors, to experience
fewer interactions. On the contrary, value of $\beta$ above zero
strengthens the positive correlation between the number of neighbors
and the total groups including him, that is, individuals with
high-degree and low-degree participate more and fewer PGGs,
respectively. The heterogeneity of actual interactions of
individuals is partly subject to change of the parameter $\beta$ and
the larger the stronger. Thus, large $\beta$ favors the emergence of
cooperation. Moreover, the monotonical dependence of degree of
cooperation on $\delta$ is quite clear for $\beta$ taking value of
the inverse of unit, and ebbs as $\beta$ goes to infinity. This is
intuitively straightforward to understand because discrepancy of the
two heterogeneity peaks at $\beta=-1$ (among all the explored cases)
and wanes for increasing $\beta$. A careful comparison shows that in
the degree-based partner selection, the number of the PGGs each
individual participates is statistically invariable no matter how
his strategy evolves. Unlikely, in another studied selection, one
cannot expect how many groups involve a given individual as the
decisive role of topology playing in participation is weakened. This
hinges in part on the different dynamics of the two selection
regimes.
\begin{figure}
\includegraphics[width=\columnwidth]{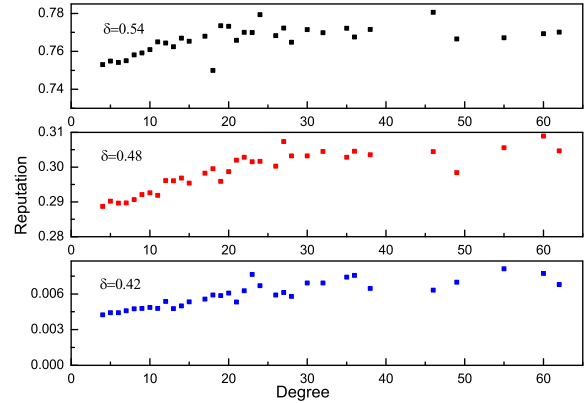}
\caption{(Color online) Reputation of individuals as a function of
degree at stationary state. A focal individual selects his $g$
co-players among his neighbors based on their reputations. Similar
method is adopted to collect data as in Fig. 2. The parameter $g$ is
also set to be $4$.}\label{fig4}
\end{figure}

The inherent complexity of games on the scale free graphs makes
analytical investigations almost impossible. But, a rough
calculation has revealed a conclusion on regular graph, which is
indeed present but not easily found on scale free graph. Without
loss of generality, let us consider the simplest case for regular
graphs with degree $k$. Assume that, for the equilibrium local
configuration, there have $k_C (k_C^{'})$ cooperators and $k_D
(k_D^{'})$ defectors around a $C (D)$. Because of the updating rule
(local competition between nearest neighbors) we used, a cooperator
on average has more cooperator neighbors than a defector. Namely,
the assortment between cooperators is induced. Hence, we have $k_C =
k_C^{'} + \triangle k$ with $\triangle k
> 0$. In addition, we have the same renormalized enhancement factor $\delta$
for different group sizes.

For group size $g = 2$, the expected average payoff of a cooperator
and a defector is given by
\[\bar{f}_C=\delta
\frac{k_C}{k}+\delta-1\]
\[\bar{f}_D=\delta
\frac{k_C^{'}}{k}\] Thus the payoff difference for $g=2$ is
$\triangle f_2=\overline{f}_C-\overline{f}_D=\delta \frac{\triangle
k}{k}+\delta-1$. Analogously, for $g=3$, the expected payoff of a
cooperator and a defector can be expressed as
\[ \bar{f}_C=\delta\left(
\frac{2k_Ck_D}{k(k-1)}+\frac{2k_C(k_C-1)}{k(k-1)}\right)+\delta-1 \]
\[\bar{f}_D=\delta\left(
\frac{k_C^{'}d_D^{'}}{k}+\frac{2k_C^{'}(k_C^{'}-1)}{k(k-1)}\right)\]
Thus the payoff difference for $g=3$ is $\triangle
f_3=\overline{f}_C-\overline{f}_D=\delta \frac{2\triangle
k}{k}+\delta-1$. Obviously, we have $\triangle f_3>\triangle f_2.$
This means that larger group size increases the payoff of
cooperators more than that of defectors. As a result, larger group
size requires a lower critical $\delta$ for the emergence of
cooperation, a well consistent analytical prediction for the
simulation experiments (see Fig. 6). Moreover, this effect exists in
any type of network no matter how heterogeneous it
is~\cite{Ohtsuki06n}.
\begin{figure}
\includegraphics[width=\columnwidth]{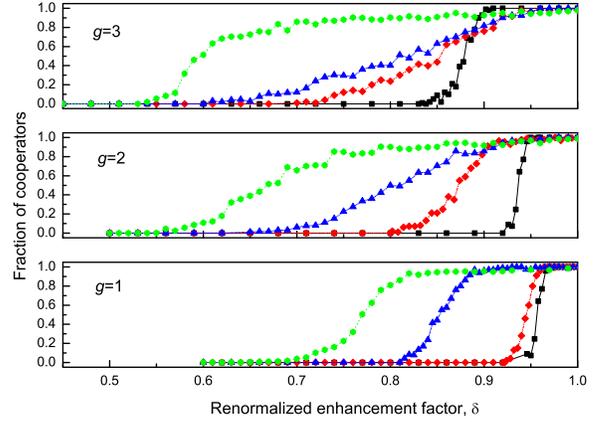}
\caption{(Color online) Fraction of cooperators as a function of the
parameter, $\delta$, with different values of the participation
weight, $\beta$. Lines with squares, diamonds, triangles, circles
correspond to the values of $\beta$ being -1, 0, 1, $\infty$.
Population structure adopted is similar to that in Fig. 1. We have
collected the data points as in Fig. 1.} \label{fig5}
\end{figure}

\begin{figure}
\includegraphics[width=\columnwidth]{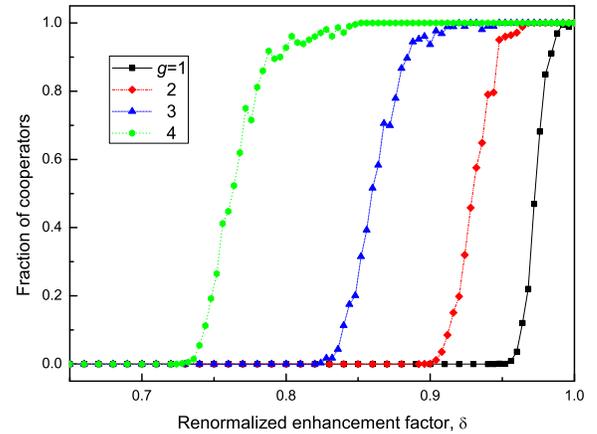}
\caption{(Color online) Fraction of cooperators as a function of the
parameter, $\delta$, with different group sizes, $g+1$. Population
structure is specified by a degree-regular graph, in which each
vertex has four neighbors. Focal individuals choose their remaining
group members due to the degree-based partner selection as in Fig.
1. Each data point is averaged over 300 runs with independent
initial strategy distributions.} \label{fig6}
\end{figure}
Taking together, although the physical networks used to specify the
population structure are independent of evolutionary courses, the
actual game interactions of focal individuals can be adjusted over
time in our minimalist model, thus leading to that the interaction
graph for a given individual does not overlap with the learning
graph being in concord with the population structure. In
Ref.~\cite{Ohtsuki07prl}, the authors have explored the evolutionary
dynamics on graphs with breaking the coincidence of the interaction
graph and the replacement graph. The approach of generating the two
subgraphs in our model is somewhat different from that in
Ref.~\cite{Ohtsuki07prl}, where difference of the two graphs can be
exactly adjusted by tuning a model parameter. In present work, as
long as the parameter $\beta$ takes finite value in the degree-based
partner selection, focal individuals each can encounter his
$\emph{g}$ neighbors probabilistically in each round. This dynamical
constitution of interacting group makes this deviation to be
dynamically changed and impossible to be precisely predicted. Then
our model can fall somewhere between games on static graphs and
coevolution of individual strategy and
neighborhood~\cite{Zimmermann04pre, Zimmermann05pre, Pacheco06prl}
in the sense that individuals not only update their strategies but
also dynamically choose group members from their neighborhood.
Noting that focal individuals propose to enter collective actions
and have choosiness towards their neighbors of different
characteristics. This is clearly an unidirectional selection but
simpler than that in Ref.~\cite{Coricelli04j} investigated
experimentally. The case of random selection (i.e., $\beta=0$)
eliminates the preferences of focal individuals towards neighbors,
but imposes a great influence on the evolution of cooperative
behavior just similar to a gradual but long lasting process of
erosion by water to the formation of deep valleys. Extension of the
random selection (i.e., non-zero values of $\beta$) closely reflects
such real-life situations where individuals have different
preferences towards individuals with diverse social ties.

In the case of participation being compulsory, an increase in the
average group size is detrimental to the survival of cooperation,
holding for both two investment schemes adopted
(see~\cite{Santos08N} for details). Conversely, our findings show
that increase in the average group size (i.e., $g+1$) is beneficial
to the buildup of cooperation on heterogeneous structured
population. In Ref.~\cite{Santos08N}, increase of the average
connectivity leads to increase in the connectedness of the graph, a
disadvantageous feature against the establishment of cooperation. In
our model, since we have fixed the average degree of the graph, the
heterogeneity of the population is unchanged. Besides, we have
normalized the enhancement factor $r$ divided by $g+1$ as $\delta$,
the overall scaling of the value of $r$ deduced from increase in
group size is automatically incorporated. Remaining differences
between curves (see Fig. 1) should be attributed to other factors.
For small fixed group size, any individual can choose $g$ group
members from his more than $g$ neighbors. This flexibility of
selection, which softens up the heterogeneity of actual interacting
network, is lessened as $g$ increases. Of an extreme $g$ being set
to be $4$, individuals accounting for a large proportion of the
population have no alternative neighbors but to confront all their
neighbors, meaning that the heterogeneity of interacting network is
much closer than that of cases of small $g$s, to the heterogeneity
of the scale free graph defining the population structure. This can
explain the rough monotonical dependence of cooperation level on the
group size $g+1$. Similarly, when we alter the participation weight,
$\beta$, increase of it also plays a striking positive role in
shrinking the divergence of the actual interacting network from the
population structure. Thus, for constant group size, cooperation is
convenient to emerge and be maintained for large $\beta$s,
consistent with simulation results (see Fig. 5).

In contrast with degree-based partner selection, the population
structure is no longer an overriding determinant on who-meets-whom
even statistically in reputation-based selection regime. Focal
individuals can unidirectionally~\cite{Coricelli04j} adjust their
preferences towards their neighbors after each round according to
their reputations, tantamount to a type of 'soft' punishment where
no cost is involved with punishers and the punished. Focal
individuals potentially exert punishment on their neighbors of ill
repute~\cite{Fehr04n, Rockenbach09n} by excluding them in the future
collective behaviors without damaging their own
reputations~\cite{Rockenbach09n}. Thus only those continuously
upholding good reputations will not suffer from being excluded and
herein acquire more opportunities of help and being helped (i.e.,
direct reciprocity). This selection regime in effect avoids 'the
second-order free-rider problem' because punishers withdraw
interactions with the punished at no cost to themselves. Thus,
cooperation should emerge for substantially small $\delta$ analogous
to degree-based regime, confirmed by the simulation results (see
Fig. 1 and Fig. 3). We finally point out that in both selection
regimes, irrespective of the social ties or reputations, no
individual can be absolutely banished in that each individual would
participate at least one PGG centered on himself.

\section{Conclusion}
We have proposed a model to study the effect of the constant group
size on the evolution of cooperation. The interactions of
individuals was metaphorized by the Public Goods Game. Two selection
regimes were introduced for a centered individual to pick up a fixed
number of players from his neighbors, according to their degrees or
reputations, which we refer to as partner selection based on degree
or partner selection based on reputation. The centered individual
adding these chosen neighbors constitutes an interacting group and
plays the PGG. Whenever individuals interact without memory effect
(only considering the social viscosity of neighbors), on the one
hand large group improves the payoff of both cooperators and
defectors, but the former at a larger rate; on the other hand, large
group strengthens the heterogeneity of the actual interactions.
These two considerations together leads to that large groups favor
cooperation more than small ones. In extended cases we found
increase of the participation weight plays a principal role in
promoting cooperation for a given group size. Later, we investigate
how cooperation evolves under the reputation-based selection regime.
The positive correlation between individual's reputation and the
number of the PGGs per individual forms a positive feedback, which
enables centered individuals to reply promptly to frequently
defective neighbors. As a consequence, cooperation can be induced to
a higher level than in the partner selection based on degree, where
individual is inept to displace the members in the community
centered on it from the viewpoint of statistics. Besides, we
scrutinize the microscopic characteristics to find that, when the
selection is progressing based on reputation, the hubs 'loyally'
play a leading role in enhancing and stabilizing cooperation in the
whole range of the renormalized enhancement factor $\delta$, which
is shortened in the degree-based selection regime, offering a better
interpretation to the aggregate observations. We also mathematically
prove that expansion in the group benefits cooperators more than
defectors in a homogeneous population, and this effect can be
generalized to any type of network, particularly scale free network.
Heterogeneity of the PGGs per individual combining the partner
selection regime together leads to rich dynamics on heterogeneous
population.

\begin{acknowledgments}
The authors are supported by NSFC (Grant Nos.~60674050, 60736022 and
60528007), National 973 Program (Grant No.~2002CB312200), National
863 Program (Grant No.~2006AA04Z258), and 11-5 project (Grant
No.~A2120061303). F. F. also gratefully acknowledges the support
from China Scholarship Council (2007U01235).
\end{acknowledgments}

\end{document}